\documentclass[%
 footinbib,
 twocolumn,
superscriptaddress,
longbibliography,
 amsmath,amssymb,
 aps,
 floatfix,
prb,
color,
nobalancelastpage,
]{revtex4-1}

\usepackage{braket}
\usepackage{tikz}
\usepackage{dsfont}
\usepackage{siunitx, xspace}
\usepackage{array}
\usepackage[version=4]{mhchem}

\newcolumntype{M}[1]{>{\centering\arraybackslash}m{#1}}

\newcommand{\eg}{{$e_g$}\xspace}
\newcommand{\tg}{{$t_{2g}$}\xspace}

 \usepackage{pgfplots}
  \pgfplotsset{compat=newest}

\usetikzlibrary{external}
\usetikzlibrary{calc,  shapes, backgrounds, arrows, chains, matrix, positioning, scopes,
                decorations, decorations.pathmorphing,patterns, fit, pgfplots.groupplots, plotmarks}
\tikzset{>=latex}
\usetikzlibrary{external}
\usepackage{graphicx}
\usepackage{dcolumn}
\usepackage{bm}

\usepackage{hyperref}

\begin{document}

\title{Dynamical Mean-Field Theory on the Real-Frequency Axis: p-d Hybridizations and Atomic Physics in SrMnO$_3$}

\author{Daniel Bauernfeind}
\email[]{daniel.bauernfeind@tugraz.at}
\affiliation{%
  Institute of Theoretical and Computational Physics\\
  Graz University of Technology, 8010 Graz, Austria
}%

\author{Robert Triebl}
\affiliation{%
  Institute of Theoretical and Computational Physics\\
  Graz University of Technology, 8010 Graz, Austria
}%

\author{Manuel Zingl}
\affiliation{%
  Institute of Theoretical and Computational Physics\\
  Graz University of Technology, 8010 Graz, Austria
}%
\affiliation{%
 Center for Computational Quantum Physics, Flatiron Institute\\
 New York, New York 10010, USA
}%

\author{Markus Aichhorn}
\affiliation{%
  Institute of Theoretical and Computational Physics\\
  Graz University of Technology, 8010 Graz, Austria
}%

\author{Hans Gerd Evertz}
\affiliation{%
  Institute of Theoretical and Computational Physics\\
  Graz University of Technology, 8010 Graz, Austria
}%
\affiliation{%
  Kavli Institute for Theoretical Physics \\ 
  University of California, Santa Barbara, CA 93106, USA
}%

\date{\today}

\begin{abstract}
We investigate the electronic structure of \ce{SrMnO3} with Density Functional Theory (DFT) plus Dynamical Mean-Field Theory (DMFT). Within this scheme the selection of the correlated subspace and the construction of the corresponding Wannier functions is a crucial step. Due to the crystal field splitting of the Mn-$3d$ orbitals and their separation from the O-2$p$ bands, \ce{SrMnO3} is a material where on first sight a 3-band $d$-only model should be sufficient. However, in the present work we demonstrate that the resulting spectrum is considerably influenced by the number of correlated orbitals and the number of bands included in the Wannier function construction. For example, in a $d$-$dp$ model we observe a splitting of the \tg lower Hubbard band into a more complex spectral structure, not observable in $d$-only models. To illustrate these high-frequency differences we employ the recently developed Fork Tensor Product State (FTPS) impurity solver, as it provides the necessary spectral resolution on the real-frequency axis. We find that the spectral structure of a 5-band $d$-$dp$ model is in good agreement with PES and XAS experiments. Our results demonstrate that the FTPS solver is capable of performing full 5-band DMFT calculations directly on the real-frequency axis. 

\end{abstract}

\maketitle

\section{Introduction}

The combination of density functional theory (DFT) and dynamical mean-field theory (DMFT) has become the work-horse 
method for the modeling of strongly-correlated materials~\cite{DFTDMFT1, Lechermann_DMFTwithWannier, 
Kotliar_ElStrCal_DMFT}. 
For DMFT, a (multi-orbital) Hubbard model is constructed in a selected correlated subspace, which usually describes the 
valence electrons of the transition metal orbitals in a material. An adequate basis for these localized orbitals are projective Wannier functions~\cite{Anisimov_Proj,TRIQS_DFTTOOLS_1}. In contrast to the Bloch wave functions, these functions are localized in real space, and therefore provide a natural basis to include local interactions as they resemble atomic orbitals and decay with increasing distance from the nuclei. However, the selection of the correlated subspace itself and the Wannier function construction are not uniquely defined.
\\
In the present work, we use \ce{SrMnO3} to analyze the differences of some common models. This perovskite is an insulator \footnote{Although most published work suggest that the compound is insulating, the experimental magnitude of the gap ranges from approximately \SI{1.0}{\electronvolt} 
to \SI{2.0}{\electronvolt}, see citations in the main text.} with a nominal filling of three electrons in the Mn $3d$ shell. There are various works concerning its electronic structure, both on the experimental~\cite{LeeExp,ABBATE_1992_T2GEGT2G,Chainani1993,KANG2008_EGT2GEG,Saitoh_EGEGT2G_minus13, 
KIM2010_EGT2GEG_minus13EV} as well as on the theoretical side~\cite{Millis5BandSrMnO3, Millis5BandSrMnO3_2,SondenaHexVSCubic,JernejSrTcO}. For the construction of the correlated subspace, we explicitly identify the following meaningful cases: The first is a three orbital model for the \tg states only. For the second choice, usually denoted as $d$-$dp$ model, the transition metal $3d$-states and the oxygen $2p$-states are considered 
in the Wannier function construction, but the Hubbard interaction is only applied to the $3d$-states. The correlated subspace is then affected by the lower lying oxygen bands due to hybridizations. In both cases, the full $3d$ manifold can be retained by including the \eg orbitals in genuine 5 orbital models.
\\
To assess the consequences of the different low-energy models, a good resolution of the spectral function on the real-frequency axis is beneficial. Due to its exactness up to statistical noise, Continuous Time Quantum Monte Carlo (CTQMC) is often used as a DMFT impurity solver~\cite{WernerMillisHybExpCTQMC, GullCTQMC, WernerFirstCTQMC}. However, when using a CTQMC impurity solver, an analytic continuation is necessary, 
which results in spectral functions with a severely limited resolution at higher frequencies~\cite{FTPS}. This can make 
it difficult to judge the influence of the choices made for the correlated subspace. In the present paper, we therefore 
employ the real-frequency Fork Tensor Product States (FTPS) solver~\cite{FTPS}.
This recently developed zero temperature impurity solver was previously applied to \ce{SrVO3}, making it possible to 
reveal an atomic multiplet structure in the upper Hubbard band~\cite{FTPS}. This observation of a distinct multiplet 
structure in a real-material calculation is an important affirmation of the atom-centered view promoted by DMFT. 
\\
The present work also serves as a deeper investigation of the capabilities of the FTPS solver. We show that the 
FTPS solver can be applied to $d$-$dp$ models, 
leading to new insight into the interplay of the atomic physics of the transition metal impurity and hybridization 
effects with the oxygen atoms as a natural extension to the atom-centered view. Furthermore, the physics of \ce{SrMnO3} 
is different from \ce{SrVO3}, since the manganate is an insulator, and thus it 
constitutes a new challenge for the FTPS solver. While we presented a proof of concept for FTPS on a simple 5-band 
model before~\cite{FTPS}, we now perform full 5-band real-frequency DFT+DMFT calculations for both 
$d$-only and $d$-$dp$ models.
\\
We find that the choices made for the correlated subspace strongly affect the resulting 
spectral function and its physical interpretation. Additionally, we show that the interplay of atomic and hybridization 
physics can already be found in very simple toy models.
\\
This paper is structured as follows. In section~\ref{sec:method} we discuss the methods employed, namely DFT, the 
different models obtained from different Wannier constructions, DMFT, and the impurity solvers used.   
Section~\ref{sec:DMFTDOS} focuses on the results of the DMFT calculations and the underlying physics of these different 
models. This knowledge will then be used in Sec.~\ref{sec:ComparisonToExp} to compare the spectral function to 
experiments by Kim \emph{et al.}~\cite{KIM2010_EGT2GEG_minus13EV}.

\section{METHOD} \label{sec:method}
\subsection{DFT and WANNIER BASIS}

\begin{figure}[t]
   \centering
   \includegraphics{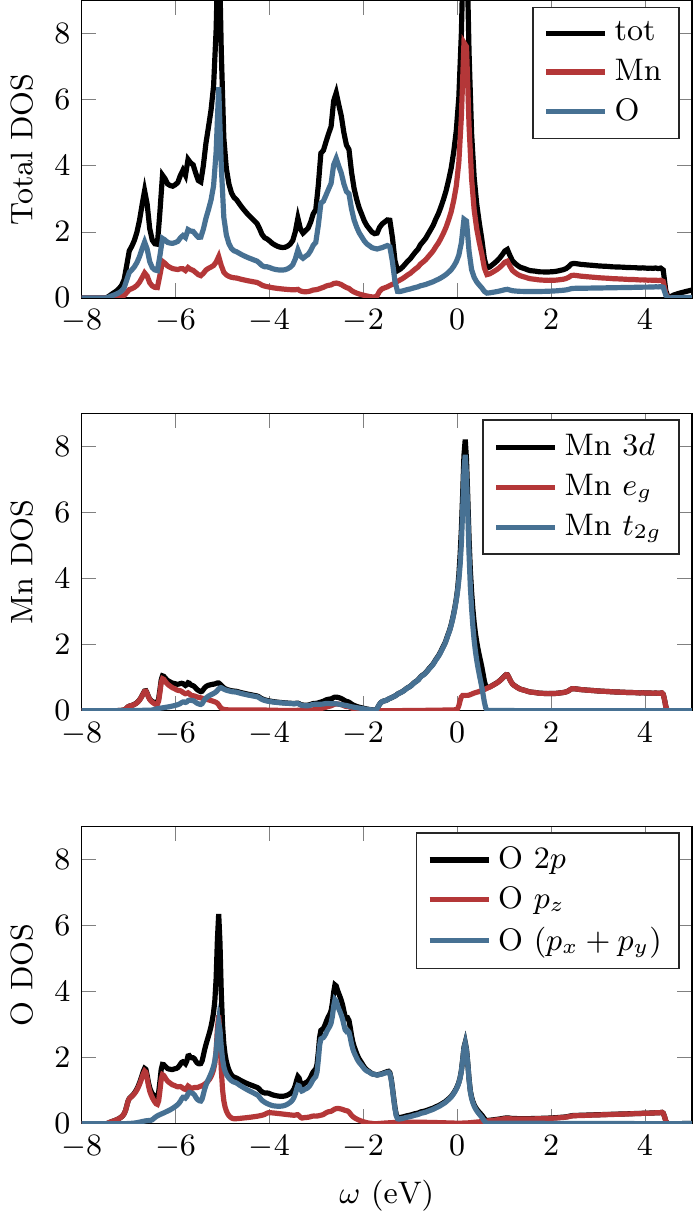}
   \caption{\emph{Top:} Total DFT-DOS for \ce{SrMnO3}. \emph{Middle:} partial Mn-3$d$ DOS. \emph{Bottom:} partial 
O-2$p$ DOS for \ce{SrMnO3}. 
   Below approximately \SI{-1.5}{\electronvolt}, the band structure consists of oxygen bands that have mostly 
$p$-character but also have some \eg and \tg weight due to hybridizations. The \tg bands are located around the Fermi 
energy from \SI{-1.5}{\electronvolt} to about \SI{0.5}{\electronvolt}, which have small $p$-character. Directly above the Fermi energy and partly overlapping with the \tg bands we find the \eg bands that have small 
$p$-contributions as well.   }
   \label{fig:DFTDOS}
\end{figure}

We start with the DFT density of states (DOS) from a non-spin-polarized DFT
calculation for \ce{SrMnO3} in the cubic phase (shown in Fig.~\ref{fig:DFTDOS}). The
calculation was performed with Wien2k~\cite{Wien2k}, using 969 $k$-points in the irreducible Brillouin zone and a
lattice parameter of $a=\SI{3.768}{\angstrom}$. Around the Fermi energy $E_F$,
\ce{SrMnO3} has the characteristic steeple-like shaped DOS, stemming from the
Mn-\tg bands with a bit of O-$p_{x/y}$ contribution. Below
\SI{-2.0}{\electronvolt}, the DOS is mainly determined by oxygen bands which
also exhibit manganese hybridizations. With the exception of some additional weight below \SI{-5.0}{\electronvolt},
the Mn-\eg states lie mainly in the energy range from \SI{0.0}{\electronvolt} to
\SI{5.0}{\electronvolt}.
\\
In this work we use projective Wannier functions, where an energy interval has to be chosen as a projection window \cite{Anisimov_Proj, TRIQS_DFTTOOLS_1}. 
The bands around $E_F$ have mainly \tg character,
suggesting a selection of only a narrow
energy window for the Wannier function construction (\SI{-2.0}{\electronvolt} to \SI{0.82}{\electronvolt}). We call 
this set of projective Wannier functions the 3-band $d$-only model. 
However, the \tg orbitals also show a considerable hybridization with the O-$2p$ states below \SI{-2}{\electronvolt}, 
and hence, one might want to enlarge the projective energy window by setting its lower boundary to 
\SI{-10}{\electronvolt}. We refer to this model as the 3-band $d$-$dp$ model. 
\\
At the same time, we realize that also the \eg orbitals are
not entirely separated from the \tg orbitals in energy and that they have even some weight around
$E_F$ (see middle graph of Fig.~\ref{fig:DFTDOS}). These states lie directly above $E_F$ and therefore their influence on the resulting
spectrum needs to be checked. One should then use a window capturing 5 bands, the \eg and \tg, as a correlated
subspace (from \SI{-2}{\electronvolt} to \SI{5}{\electronvolt}). This is a
5-band $d$-only model. Note that empty orbitals do not pose a problem for the FTPS solver. Like before, we can again 
enlarge the energy window to include the oxygen hybridization (\SI{-10}{\electronvolt} to
\SI{5}{\electronvolt}). We denote this model as the 5-band $d$-$dp$ model. 
\\
In total, we end up with 4 different choices. The settings for these 4 models are summarized
in Tab.~\ref{tab:models}. All of them are justified, have different
descriptive power, and have been employed in various DFT+DMFT
calculations for \ce{SrMnO3}~\cite{Millis5BandSrMnO3_2,Millis5BandSrMnO3,JernejSrTcO}.

\begin{table*}[t]
\setlength\extrarowheight{8pt}
\caption{Summary of models with their projective energy
windows and the parameters used in the FTPS solver: number of bath
sites $N_B$, Fourier transform broadening $\eta_{FT}$, truncated weight
$t_w$ and maximal bond dimension allowed for the links between impurities as well as
for the links between an impurity and the first bath tensor. We keep at most 
this number of states and increase $t_w$ if needed. The number in
brackets is the maximal bond dimension during ground state search, while
the first number is used for the time evolution. The bath links were not
restricted to any maximal bond dimension. The FTPS time evolution is performed up to $t_{max}$, given in eV$^{-1}$. }
  \label{tab:models}
  \begin{ruledtabular}
  \centering
  \begin{tabular}{ ccc|ccccc }
  \centering

      \textbf{Model} & \textbf{Window ($eV$)} & \textbf{Comments} & $\mathbf{
N_B } $ & $\bm{ \eta_{FT} } $ & $\mathbf{ t_w } $ & \textbf{Bond dim.}  & $\bm{t_{max}}$
\\ \hline

    3-band $d$-only & $-2.0$ - $0.82$ & only major \tg weight around $E_F$ & 79 &
0.08 & $5\cdot10^{-9}$ & - & 14.0 \\ \hline

    5-band $d$-only & $-2.0$ - $5.0$ & include \eg, neglect hybridizations & 49
 & 0.15 & $1\cdot10^{-8}$ & 200 (150) & 12.0  \\
\hline

    3-band $d$-$dp$-model &$-10.0$ - $5.0$ & include hybridized \tg weight on
 oxygen bands & 59 & 0.1 &
$1\cdot10^{-8}$ & 450 (150) & 14.0  \\ \hline

    5-band $d$-$dp$-model &$-10.0$ - $5.0$ & \tg and \eg bands with
hybridizations & 49 & 0.2 & $1\cdot10^{-8}$ & 200 (150) & 7.0 \\

   \end{tabular}
  \end{ruledtabular}
\end{table*}

\subsection{DMFT}
Once the correlated subspace is defined, we use DMFT~\cite{GeorgesDMFT, MetznerVollhardt_Dinf,Lechermann_DMFTwithWannier, Kotliar_ElStrCal_DMFT} to solve the resulting multi-band Hubbard model. As interaction term we choose the 5/3-band Kanamori Hamiltonian~\cite{KanamoriH}\footnote{Even for the five-band calculations we choose the Kanamori Hamiltonian over the Slater Hamiltonian~\cite{Slater}, because the large number of interaction terms in the latter make a treatment with the FTPS solver more involved.}. Within DMFT, the lattice problem is mapped self consistently onto an Anderson impurity model 
(AIM) with the Hamiltonian

 \begin{align}\label{eq:H_DMFT3B}
	H &=   H_{\text{loc}} + H_{\text{bath}}  \\
	H_{\text{loc}} &=  \sum_{m \sigma}\epsilon_{m0} n_{m 0 \sigma} + H_{
\text{DD} } + H_{\text{SF-PH} }\nonumber \\
	H_{\text{DD}} &=  U \sum_m n_{m0\uparrow} n_{m0\downarrow}
        \nonumber \\
        &+(U-2J) \sum_{ m'>m, \sigma } n_{m0\sigma} n_{m' 0\bar{\sigma}}
\nonumber\\
	  &+(U-3J) \sum_{ m'>m, \sigma } n_{m0\sigma} n_{m' 0 \sigma} \nonumber
\\
	H_{ \text{SF-PH} } &= J\sum_{m'>m} \left ( c_{m 0 \uparrow
          }^{\dag} c_{m 0 \downarrow }    c_{m' 0 \uparrow } c_{m' 0
            \downarrow }^{\dag} +  \text{h.c.} \right ) \nonumber\\
	 	  &-J\sum_{m'>m} \left( c_{m 0 \uparrow }^{\dag} c_{m 0
\downarrow }^{\dag} c_{m' 0 \uparrow } c_{m' 0 \downarrow } + \text{h.c.}
\right )  \nonumber\\
 	H_{ \text{bath} } &= \sum_{m l \sigma} \epsilon_{ml} n_{ml\sigma} + V_{ml}
\left ( c_{m0\sigma}^{\dag} c_{ml\sigma} + \text{h.c.} \right ).\nonumber
\end{align}
Here, $c_{m l \sigma}^{\dag}$ ($c_{m l \sigma}$) creates (annihilates) an
electron in orbital $m$, with spin $\sigma$ at site $l$ (site zero is the
impurity). $n_{m l \sigma}$ are the corresponding particle number operators. $\epsilon_{m0}$ is the orbital dependent 
on-site energy of the impurity and $\epsilon_{ml}$ as well as $V_{ml}$ are the bath on-site energies and the 
bath-impurity hybridizations, respectively.
\\
The interaction part of Hamiltonian~\eqref{eq:H_DMFT3B}, $H_{\text{DD}} + H_{\text{SF-PH} }$, is parametrized by a 
repulsive interaction $U$ and the Hund's coupling $J$. For each of the models presented in Tab.~\ref{tab:models}, we 
choose these parameters \emph{ad hoc} in order to obtain qualitatively reasonable results. In addition, for the full 
5-band $d$-$dp$ model we also estimate them quantitatively via a comparison to an experiment. 
\\
Within DFT+DMFT, a so-called double counting (DC) correction is necessary, because part of the electronic correlations 
are already accounted for by DFT. For general cases, exact expressions for the DC are not known, although there exist 
several approximations~\cite{HauleExactDC,ParkMillis_UprimeDC,karolakLichtenstein_DC,Haule_DCstatic}. In the present 
work we use the fully-localized-limit (FLL) DC (Eq.(45) in Ref.~\onlinecite{FLL_Held}). When needed, we adjust it 
to account for deviations from the true, unknown DC. Note that in the $d$-only models, the DC is a trivial energy shift 
that can be absorbed into the chemical potential~\cite{karolakLichtenstein_DC}, which is already adjusted to obtain the 
correct number of electrons in the Brillouin zone. This step, as well as all other interfacing between DFT and DMFT, is 
performed using the TRIQS/DFTTools package (v1.4)~\cite{TRIQS, TRIQS_DFTTOOLS,TRIQS_DFTTOOLS_1,TRIQS_DFTTOOLS_2}.

\subsection{CTQMC + MaxEnt}
We compare some of our results to CTQMC data at an inverse temperature of
$\beta=\SI{40}{\electronvolt^{-1}}$ obtained with the TRIQS/CTHYB solver (v1.4)~\cite{TRIQS_CTHYB, WernerMillisHybExpCTQMC}. We calculate real-frequency spectra with an analytic continuation using the freely available $\Omega$-MaxEnt implementation of the Maximum Entropy (MaxEnt) method~\cite{OMEGA_MaxEnt}. However, the analytic continuation fails to reproduce high-energy structure in the spectral function, as we have shown in Ref.~\cite{FTPS} on the example of \ce{SrVO3}. This is especially true when the imaginary-time Green's function is subject to statistical noise, which is inherent to Monte Carlo methods. 
\\
 There are in general two quantities for which one can perform the analytic continuation. First, one can directly calculate the real-frequency impurity Green's function from its imaginary time counterpart (as done in Fig.~\ref{fig:5Band_DP}). Second, one can perform the continuation on the level of the impurity self-energy \cite{Mravlje_Thermopower} and then calculate the local Green's function of the lattice model (as done in Fig.~\ref{fig:5Band_DP_Gloc}). In the latter case, the DFT band-structure enters on the real-frequency axis, which increases the resolution of the spectral function.  

\subsection{FTPS}
For all models studied we employ FTPS~\cite{FTPS}. This recently developed impurity solver uses a tensor 
network geometry which is especially suited for AIMs. The first step of this temperature $T=0$ method is to find
the absolute ground state including all particle number sectors with DMRG~\cite{WhiteDMRG}. Then
the interacting impurity Green's function is calculated by real-time evolution. Since entanglement growth during time 
evolution prohibits access to arbitrary long times~\cite{SchollwoeckDMRG_MPS}, we calculate the Green's function up to 
some finite time (see Tab.~\ref{tab:models}) and predict the time series using the linear prediction 
method~\cite{WhiteLinPred,FTPS} up to times $\mathcal{O}(100eV^{-1})$.
The linear prediction could potentially produce artifacts in the spectrum, and therefore we always make sure that every spectral 
feature discussed in this work is already present in the finite-time Green's function without linear prediction.
\\
The main approximations that influence the result of the FTPS solver are the
broadening $\eta_{FT}$ used in the Fourier transform \footnote{We Fourier
transform with a kernel $e^{i\omega t - \eta_{FT} |t|}$ }, and the truncation of the
tensor network~\cite{FTPS}. The former corresponds to a convolution with
a Lorentzian in frequency space making its influence predictable, while the truncation can be controlled by including more states. This control over the approximations allows us to analyze spectral functions in greater detail than what 
would be possible with CTQMC+MaxEnt. The parameter values for our FTPS calculations are listed in 
Tab.~\ref{tab:models}. 
\\
Note that we choose $\eta_{FT}$ larger than in our previous work~\cite{FTPS}. The 
reason for doing this is two-fold: First, some of the calculations we show in this work have a large bandwidth, which 
lowers the energy resolution if we keep the number of bath sites fixed. Second, FTPS uses a discretized bath to 
represent the continuous non-interacting lattice Green's function $G_0^{\text{cont} }$. When calculating the self 
energy $\Sigma = G_0^{-1} - G^{-1}$, we can either 
use the discretized version of $G_0^{\text{discr} }$ or the continuous one, $G_0^{\text{cont} }$. 
In this work we choose $G_0^{\text{discr} }$, which is formally the correct choice. This 
then requires to use a larger broadening to obtain causal self-energies that do not show finite discretization effects 
from inverting $G_0^{\text{discr} }$. However, when calculating the final impurity spectral function shown in all 
figures, we employ a very small broadening of $\eta_{FT} = \SI{0.01}{\electronvolt}$ in order to obtain optimal 
resolution.
\\
The real-frequency approach of FTPS allows to resolve spectral features with higher precision than CTQMC+MaxEnt. 
This is especially true for high energy multiplets. On the other hand, with FTPS and real-time evolution it 
is difficult to obtain perfect gaps, since the results are less precise at small $\omega$, encoded in the long-time properties of the Green's function which we obtain only approximately using linear 
prediction~\cite{WhiteLinPred}. 
\\
With FTPS we calculate the greater and lesser Green's functions separately~\cite{FTPS}. Since the greater (lesser) 
Green's function has no contribution at $\omega<0$ ($\omega>0$) we restricted the contributions of the calculated Green's 
functions in frequency space.

\section{RESULTS}\label{sec:DMFTDOS}

\begin{figure}[t]
   \centering
   \includegraphics{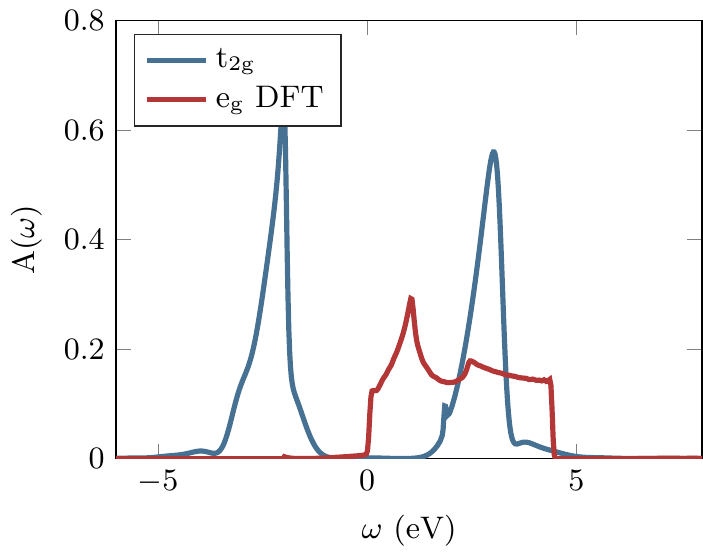}
   \caption{3-band $d$-only calculation: \tg correlated spectral function for $U = \SI{4.0}{\electronvolt}$
and $J = \SI{0.6}{\electronvolt}$, as well as \eg DFT-DOS. The \tg spectrum shows
a Mott insulator at half-filling with pronounced lower and upper Hubbard
bands.}
   \label{fig:3Band}
  \end{figure}

  \begin{figure}[t]
   \centering
   \includegraphics{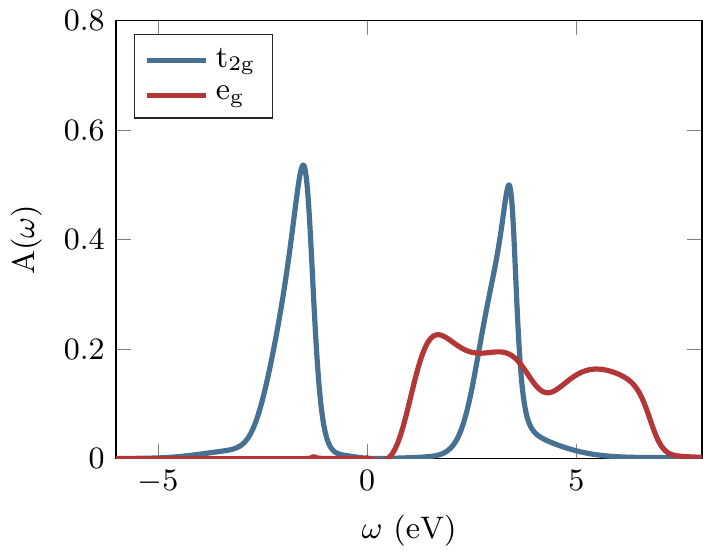}
   \caption{5-band $d$-only calculation: correlated spectral function of the \eg and the \tg orbitals for $U=$ 
\SI{4.0}{\electronvolt} and $J=$ \SI{0.6}{\electronvolt}.}
   \label{fig:5Band}
  \end{figure}

\subsection{\emph{d}-only models}
First we focus on $d$-only calculations using a projective energy window with a 
lower energy boundary of \SI{-2.0}{\electronvolt} for the Wannier-function 
construction, neglecting the occupied Mn-$3d$ weight at lower energies (see Tab.~\ref{tab:models} and middle graph of 
Fig.~\ref{fig:DFTDOS}). With this choice of the correlated 
subspace, the occupation of the \eg orbitals is nearly zero and the three 
degenerate \tg orbitals are half-filled.

\subsubsection*{3-band calculation}
Considering only the \tg subspace, the resulting impurity spectral function 
(Fig.~\ref{fig:3Band}) is gapped for the chosen interaction values. The peaks 
of the lower and upper Hubbard bands are separated by \SI{5.0}{\electronvolt} 
in energy, which is roughly $U+2J = \SI{5.2}{\electronvolt}$, as expected from atomic physics~\cite{GeorgesHundJ}.
\\
Contrary to \ce{SrVO3}, where a distinct 3-peak multiplet structure in the 
upper Hubbard band is present~\cite{FTPS}, both \ce{SrMnO3} Hubbard bands show
only one dominant peak. The structure observed in \ce{SrVO3} was well explained 
by the atomic multiplets of the interaction Hamiltonian $H_{ \text{loc} }$ in 
Eq.~\ref{eq:H_DMFT3B} for a ground state with one electron occupying the \tg 
orbitals. The absence of such an atomic multiplet structure in this model for \ce{SrMnO3} can be understood in a similar way: The 
large Coulomb repulsion in combination with Hund's rules (due to the density-density 
interaction strengths $U$, $U-2J$ and $U-3J$) lead to a ground state $\ket{\psi_0}$ which consists 
mostly of the states $\ket{\uparrow, \uparrow,\uparrow}$ and 
$\ket{\downarrow,\downarrow,\downarrow}$ on the impurity. Adding a
particle, when calculating the Green's function, produces a single double 
occupation, e.g., $c_{1,\downarrow}^{\dag}\ket{\psi_0} = \ket{\uparrow \downarrow,
\uparrow,\uparrow}$. This state is an eigenstate of the atomic Hamiltonian, because it is trivially an eigenstate of 
$H_{\text{DD}}$, and both the spin-flip and pair-hopping terms annihilate this state. Hence, all \tg single-particle excitations from the ground state have the same energy, and as a consequence, only \emph{one} atomic excitation energy is observed.
\\
Although not included in the low-energy model, the uncorrelated states still need to be taken into account for the 
single-particle gap of \ce{SrMnO3}. On the unoccupied side, the onset of the \eg orbitals leads 
to a reduction of the single-particle gap to about half the size of the \tg 
gap (see Fig.~\ref{fig:3Band}). On the occupied side, depending on $U$ and $J$, 
either the lower Hubbard band or the O-bands (at about 
\SI{-1.5}{\electronvolt}) determine the gap size, and thus also the type of the insulating state (Mott or charge transfer insulator~\cite{ZaanenChargeTransferInsulator}). For \ce{SrMnO3} to be clearly classified as Mott insulator, $U+2J < \SI{3.0}{\electronvolt}$ would be required. However, it is questionable if the $d$-only picture is correct, 
as in this case the lower Hubbard band is not influenced by the \tg/O-2$p$ hybridizations between \SI{-6.0}{\electronvolt} and 
\SI{-2.0}{\electronvolt} (see Fig~\ref{fig:DFTDOS}). We will discuss the effect of these hybridizations in detail
in Sec.~\ref{sec_3banddp} and Sec.~\ref{sec_5banddp}.

\subsubsection*{5-band calculation}
Next, we add the \eg orbitals to the correlated subspace, which now comprises the full Mn-$3d$ manifold. The resulting 
impurity spectral functions of the \eg and 
\tg orbitals are shown in Fig.~\ref{fig:5Band}.
The \tg spectral weight does not change much compared to the 
3-band calculation. This is to be expected, because the \eg orbitals remain nearly empty during the calculation of the 
\tg Green's function.
\\
The \eg spectral function, on the other hand, becomes much broader in comparison to the DFT-DOS, showing spectral weight above \SI{4.5}{\electronvolt}. The unoccupied part of the spectrum is encoded in the greater Green's function, i.e., adding a particle in an \eg orbital to the ground state. If we again assume 
$\ket{\psi_0}\propto\ket{\uparrow, \uparrow,\uparrow} + \ket{\downarrow,\downarrow,\downarrow}$ as the \tg ground 
state, we can
add a particle to the \eg orbitals either in a high-spin or low-spin
configuration:
\begin{align} \label{eq:egMinorityMajorityExcitation}
 c_{e_g\uparrow}^\dag \ket{\psi_0} \propto \underbrace{ \ket{\uparrow, \uparrow,\uparrow} }_{t_{2g}} \otimes 
\underbrace{  \ket{\uparrow,0} }_{e_g} + \ket{\downarrow, \downarrow,\downarrow} \otimes \ket{\uparrow,0}.
\end{align}
Using the Kanamori Hamiltonian, the high-spin configuration (first term in Eq.~\ref{eq:egMinorityMajorityExcitation}) 
generates a single atomic excitation energy, while the low-spin configuration (second term in 
Eq.~\ref{eq:egMinorityMajorityExcitation}) leads to two energies (due to the spin-flip terms). According to this 
atomistic picture, the splitting of the \eg peaks is proportional to Hund's coupling $J$ (see Fig.~\ref{fig:5Band}). 
Their position relative to the upper \tg Hubbard band is influenced by the crystal field splitting and $J$. From this 
clear atomic-like structure we see that even empty orbitals need to be included in the correlated subspace because of 
correlation effects with other occupied orbitals.

\subsection{3-band \emph{d}-\emph{dp} model}\label{sec_3banddp}
\begin{figure}[t]
   \centering
   \includegraphics{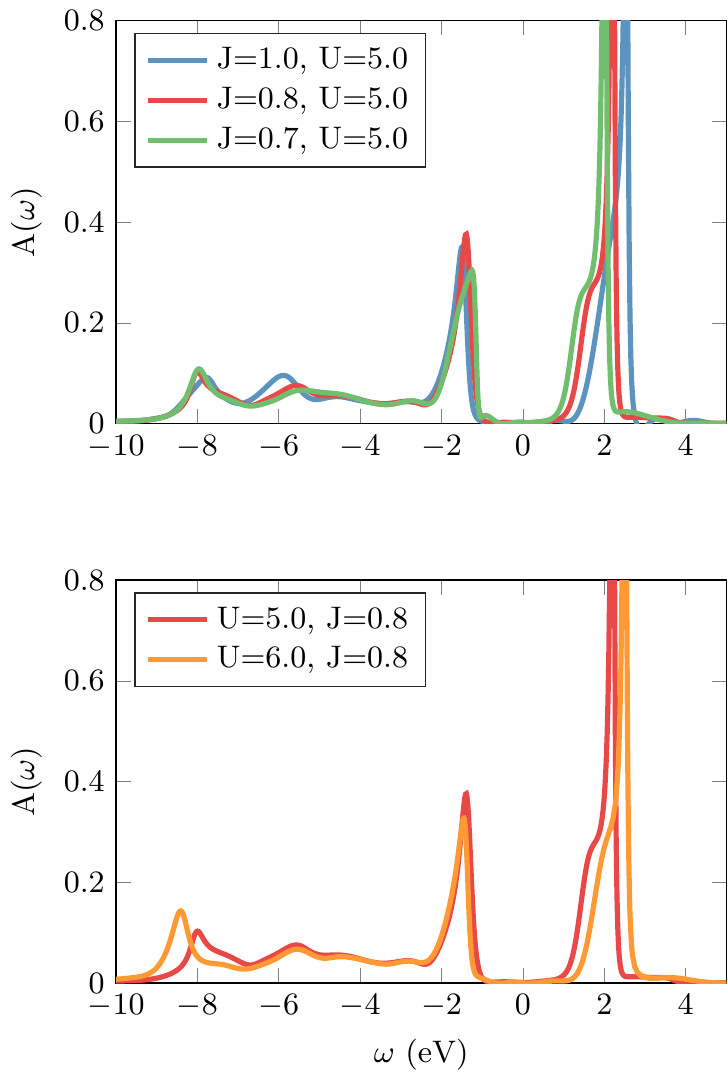}
   \caption{ 3-band \emph{d-dp} model: spectral functions for different $J$ (top) and different $U$
(bottom). All interaction parameters are given in eV. Upon increasing both parameters the gap increases. Changing $J$ 
shifts the peak at around \SI{-6.0}{\electronvolt}, while changing $U$ only
shifts the one at \SI{-8.0}{\electronvolt}. }
   \label{fig:3BandDP}
  \end{figure}
In the energy region where the lower Hubbard band is located, we also find \tg
weight stemming from the Mn-$3d$/O-$2p$ hybridization (see middle plot of Fig.~\ref{fig:DFTDOS}). This suggests that 
those states should be included in the construction of the projective Wannier functions, i.e., a $d$-$dp$ model. In the 
following we will use the term High Energy Spectral Weight (HESW) to denote the Wannier 
function weight on the oxygen bands (located below \SI{-1.5}{\electronvolt}). The first and most obvious consequence of 
a larger projective energy window is an 
increased bandwidth of the Wannier DOS. To obtain a similar insulating behavior as in the $d$-only model we 
increase $U$ and $J$. Secondly, now also the DC correction has a non-trivial effect, since it shifts the correlated \tg 
states relative to the oxygen bands. The \tg weight on the oxygen bands is 
rather small, which means that the effect of the DC correction on the HESW is equally low. 
Thirdly, in the 3-band $d$-$dp$ model the impurity occupation grows (the exact value depending on $U$ and $J$), changing 
the character of the ground state to a mix of states with mainly three and four particles on the impurity, while in the 
3-band $d$-only calculation the occupation of the impurity was three electrons. Due to the increased complexity of the 
ground state, we expect a richer dependence of the spectrum on the interaction parameters $U$ and $J$.
\\
\begin{figure}[t]
   \centering
   \includegraphics{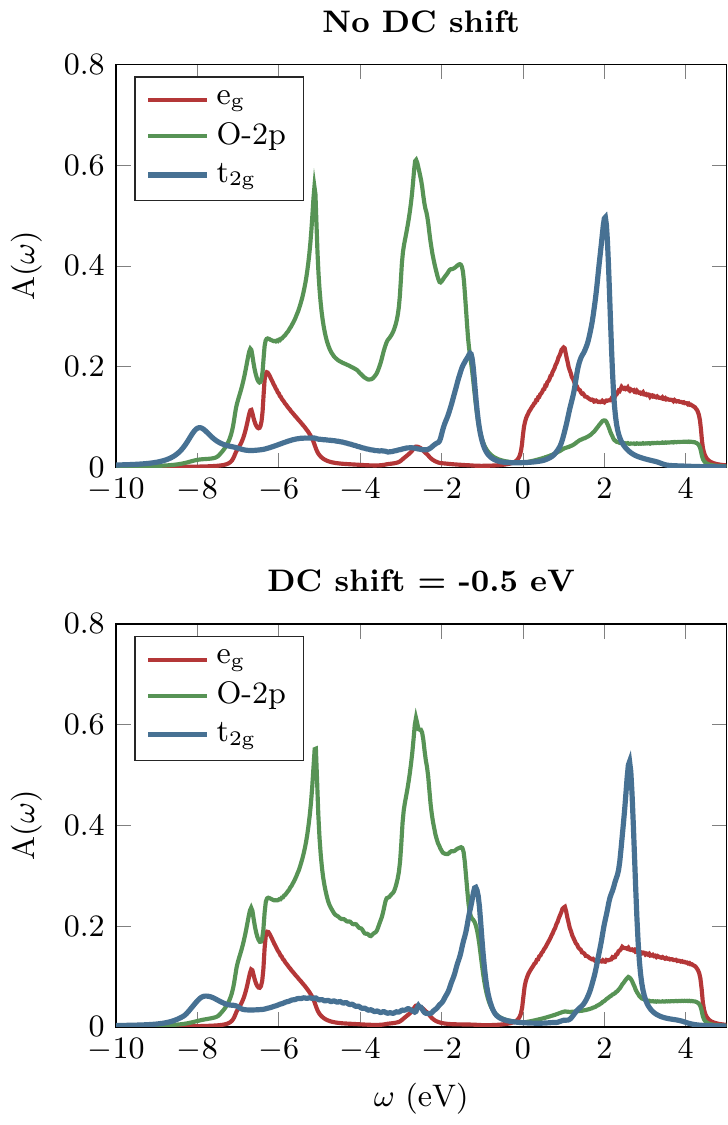}
   \caption{ 3-band \emph{d-dp} model: effect of the DC correction. \emph{Top:} FLL DC, \emph{Bottom:} FLL DC with an additional shift of \SI{-0.5}{\electronvolt}. $U=$ \SI{5.0}{\electronvolt} and $J=$ \SI{0.7}{\electronvolt} are used as interaction parameters. Contrary to all other figures, in this plot we show the spectrum of the correlated local Green's function. The occupation of the \tg orbitals changes from $0.55$ (No DC shift) to $0.54$ (DC shift =\SI{-0.5}{\electronvolt}) indicating that the \tg occupation is not much affected by the DC. The DFT-DOS (Fig.~\ref{fig:DFTDOS}) already gives an occupation of $0.54$. }
   \label{fig:3BandDP_DC}
\end{figure}
In Fig.~\ref{fig:3BandDP} we compare calculations for different values of
$J$ (top) and different values of $U$ (bottom). Overall, the spectral functions consist of a (smaller) lower Hubbard band connected to states from the hybridized oxygen bands and an upper steeple-like Hubbard band of similar shape as in the $d$-only calculation. By comparing the two peaks at \SI{-6.0}{\electronvolt} and \SI{-8.0}{\electronvolt}, we observe that they behave differently when changing $U$ or $J$. While the former is only affected by $J$, the latter is not, but shifts with $U$. The resolution of the structure in the lower-Hubbard-band/HESW complex demonstrates the capabilities of the FTPS solver.
\\
The \tg gap grows when increasing either $U$ or $J$, which is a typical sign of 
Mott physics at half filling~\cite{GeorgesHundJ}. Nevertheless, in the $d$-$dp$ model the gap size increases slowly: when increasing $U$ by \SI{1.0}{\electronvolt}, the 
gap only grows by about half of that. Considering also the uncorrelated \eg orbitals, we observe that the single-particle gap is not much affected by the interaction values studied. An artificial lowering of the DC correction by 
\SI{-0.5}{\electronvolt}, which corresponds to a relative shift in energy between the correlated subspace and the uncorrelated states, also increases the \tg gap (Fig.~\ref{fig:3BandDP_DC}). This 
growth of the gap is mostly due to a shift of the \tg upper Hubbard band, since the chemical potential is pinned by the \eg bands~\footnote{We found that the \eg bands have a very small occupation already in the DFT calculation. This does not change during the DMFT calculations and therefore, the chemical potential is always pinned at the onset of the \eg spectral function.}. The first excitation below $E_F$ has a mix of \tg and O-$p$ character. This indicates that in this model, 
\ce{SrMnO3} is not a pure Mott insulator, but a mixture between Mott- and charge transfer insulator. This classification 
is consistent with previous results~\cite{Chainani1993,Saitoh_EGEGT2G_minus13,ABBATE_1992_T2GEGT2G,Millis5BandSrMnO3}.
\\
\begin{figure}[t]
   \centering
   \includegraphics{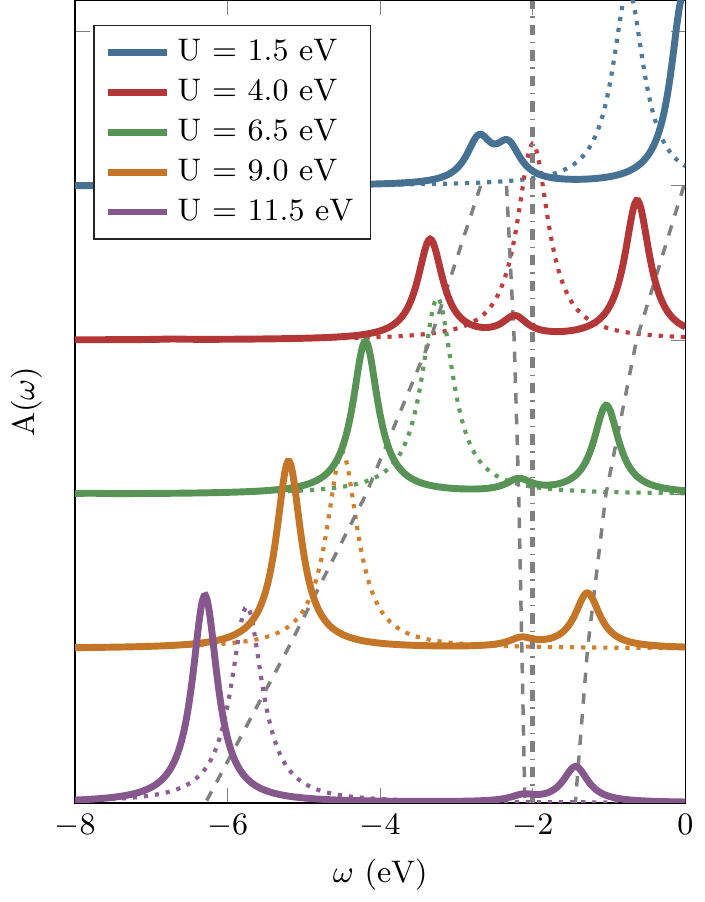}
   \caption{Spectrum of a one-band AIM with one interacting site coupled to a single non-interacting 
site (Eq.~\ref{eq:toymodel}). The spectrum is 
calculated with the absolute ground state over all particle number sectors. The on-site energy
$\epsilon_1$ is shown as gray dashed-dotted vertical line. The gray
dashed lines visualize the evolution of the location of the three peaks as a function of the
interaction strength $U$. The colored dotted peaks show the atomic
spectrum with peaks at $-U/2$. The upper Hubbard band (additional peak at $\omega>0$) is not shown. All spectra have 
been broadened by $\eta_{FT}=\SI{0.2}{\electronvolt}$.}
   \label{fig:AIM_1orb1site}
\end{figure}
Let us employ a simple toy model to qualitatively understand this intermediate regime. We use a correlated site coupled 
to only one non-interacting site:
\begin{align}
 H = &U( n_{0,\uparrow} - 0.5 ) ( n_{0,\downarrow}-0.5) \nonumber\\
 &+ \sum_\sigma V_1
(c_{0,\sigma}^\dag c_{1,\sigma} + h.c.) + \epsilon_1 n_{1,\sigma}\label{eq:toymodel}
\end{align}
The purpose of the non-interacting site is to mimic the effect of the HESW. 
We set the on-site energy to $\epsilon_1=\SI{-2.0}{\electronvolt}$ and use a coupling to the impurity of
$V_1=\SI{1.0}{\electronvolt}$. Since we want to understand the occupied
part of the spectrum, we focus on negative energies only.
In Fig.~\ref{fig:AIM_1orb1site} we show the resulting spectral functions ($\omega<0$) for
various values of the interaction strength $U$ (full lines). The atomic excitation spectra of this model (corresponding to $V_1=0$), whose peaks are positioned exactly at $-U/2$, are indicated by dotted lines. This toy model shows three important features:\\
\emph{First:} The peak highest in energy (above $\SI{-2.0}{\electronvolt}$) corresponds to the lower Hubbard 
band for small values of $U$~\footnote{If we use a larger bath energy $\epsilon_1$, for example $\epsilon_1 
=\SI{-5.0}{\electronvolt}$, the position of the first peak of the impurity spectrum is proportional to $U$ at small 
$U$, showing that it is indeed a lower Hubbard band. }. We see that it does not cross the on-site energy $\epsilon_1$ 
with increasing $U$, but approaches it asymptotically. The bath
site \emph{repels} this level and upon increasing $U$ its weight decreases.  \\
\emph{Second:} The peak lowest in energy shows the opposite behavior. The
uncorrelated site repels it towards lower energies and the spectral weight increases
when we increase $U$. For large $U$ this level asymptotically approaches the atomic limit at energy
$-U/2$ and eventually becomes the lower Hubbard band. These two peaks together form what one could call a split lower 
Hubbard band. 
\\
\emph{Third:} The excitation at the on-site energy $\epsilon_1$ shifts to lower energy and splits under the influence 
of $U$. Upon increasing $U$, one part develops into the lower Hubbard band discussed above, and the other approaches 
$\epsilon_1$ from below, with diminishing weight. \\
The DMFT spectral functions (Fig.~\ref{fig:3BandDP}) also show roughly a 3-peak 
structure, where the peaks at about \SI{-1.5}{\electronvolt} (\SI{-8.0}{\electronvolt}) could be the first 
(last) peak of the split lower Hubbard band of our toy model. The region in between then corresponds to the small, 
middle peak in the toy model stemming from  the HESW.
\\
The repulsion of the first peak explains why increasing $U$ (Fig.~\ref{fig:3BandDP} lower graph) has only a relatively 
weak effect on the size of the gap. On the other hand, effectively shifting the oxygen bands with the DC correction to 
lower energies (Fig.~\ref{fig:3BandDP_DC}) corresponds to shifting the bath energy $\epsilon_1$. This means that the 
repulsion gets weaker, which explains the growth of the gap. Furthermore, when increasing $U$ we find that the peak 
highest in energy gets smaller, while spectral weight is transferred to the lowest energy peak, which is also shifted 
to lower energies (Fig.~\ref{fig:3BandDP}). Additionally, a lowering of the DC correction leads to an opposite behavior, 
where the first peak below $E_F$ grows at the expense of the lowest one in energy. Note that the middle region of our DMFT 
spectrum shows a $J$-dependence (Fig.~\ref{fig:3BandDP} top), which cannot be explained by a one-orbital toy model. Using a similar toy model with two orbitals and Kanamori interaction, we indeed observe a splitting proportional to $J$ in the spectra (not shown here). Since the effect is small we will refrain from discussing it in more depth. 
\\
We emphasize that the close relation between the toy model and the actual impurity Green's function of \ce{SrMnO3} in the $d$-$dp$ model suggests that the HESW has the effect of splitting the lower Hubbard band into two bands; their separation 
increases with the hybridization strength. Therefore, including the oxygen states in the model strongly influences the 
size of the gap.

\subsection{5-band \emph{d}-\emph{dp}  model }\label{sec_5banddp}

\begin{figure}[t]
   \centering
   \includegraphics{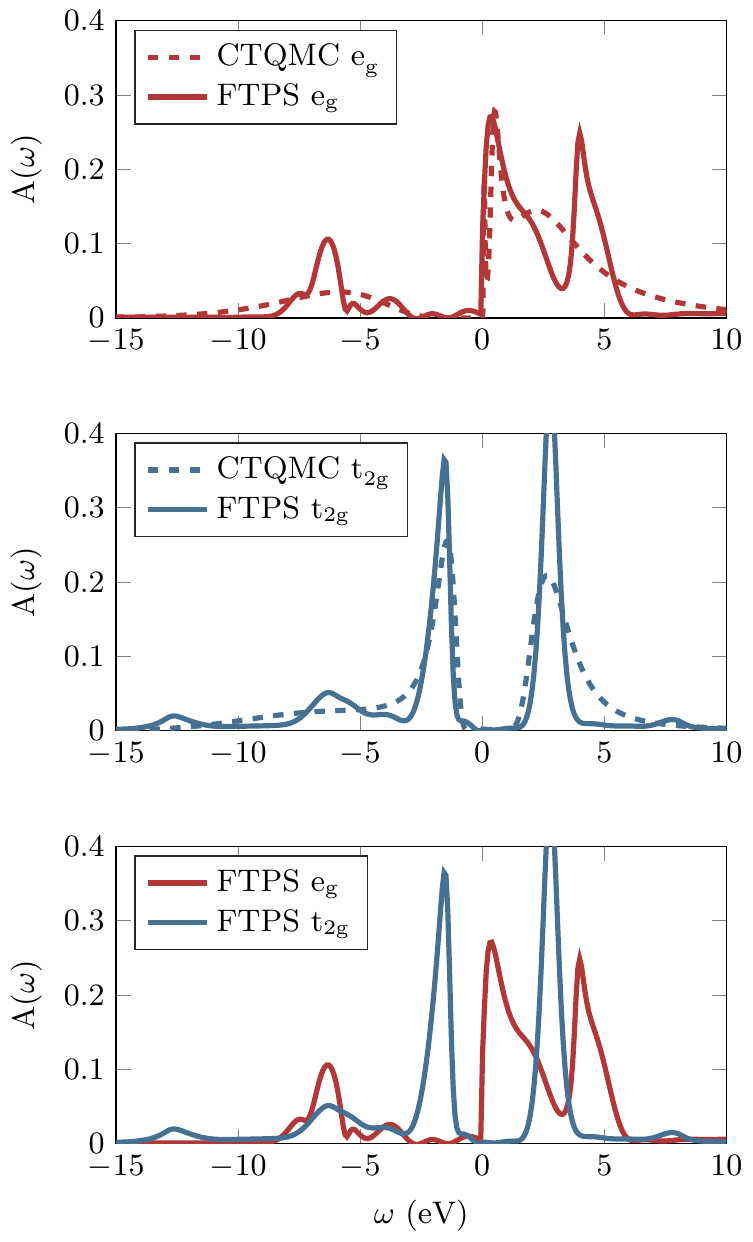}
   \caption{ Comparison of the spectral functions for the 5-band \emph{d-dp} model between FTPS and CTQMC+MaxEnt using the impurity Green's function. \emph{Top:} \eg orbitals. \emph{Middle:} \tg orbitals. For both calculations we use $U=$ \SI{6.0}{\electronvolt} and $J=$ \SI{0.8}{\electronvolt}. \emph{Bottom:} 
Combined spectral function. }
   \label{fig:5Band_DP}
  \end{figure}

  \begin{figure}[t]
    \centering
    \includegraphics{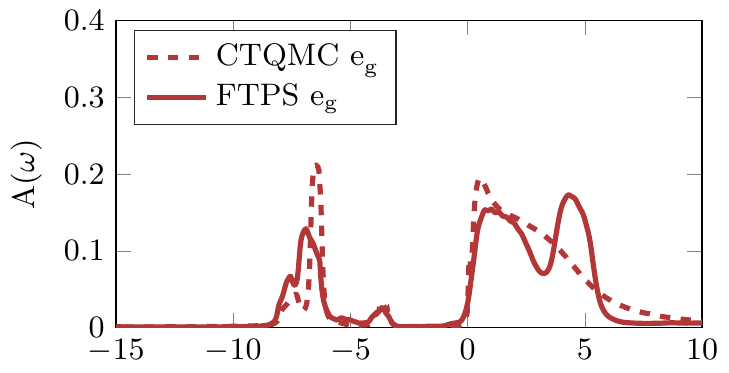}
    \caption{Comparison of the \eg spectral functions of the local lattice Green's function for the 5-band \emph{d-dp} model between FTPS and CTQMC+MaxEnt, when the analytic continuation is performed for the self-energy. To obtain the plot with FTPS we used $\eta_{FT} = 0.25$. Using the band structure on the real-frequency axis enhances the resolution of the CTQMC result at energies where the imaginary part of the self-energy is small, and increases its agreement with FTPS in the occupied part of the spectral function. }
    \label{fig:5Band_DP_Gloc}
   \end{figure}

From the DFT-DOS in Fig.~\ref{fig:DFTDOS}, we see that the \eg orbitals
are actually not empty. They possess additional spectral weight at around 
\SI{-7.0}{\electronvolt}, stemming from hybridizations with the oxygen bands. 
Similarly to the previous section where we included hybridizations of \tg and O-$2p$, we now also include the hybridizations of \eg and O-$2p$. 
\\
As mentioned at the beginning, only approximations to the DC correction are known. For the present 5-band calculation we find that using the FLL DC does not produce a pronounced gap. This can be traced back to the additional hybridizations of \eg with O-$2p$ (see discussion below). Furthermore, the FLL formula is based on five degenerate orbitals. In the case at hand we find an approximately half filled \tg impurity ($\langle n_{t_{2g} 0\sigma} \rangle \approx 0.5$) and about one electron in total on the \eg part of the impurity ($\langle n_{e_g0\sigma} \rangle \approx 0.2$). One therefore needs to adapt the DC correction to reproduce experimental results. In order to obtain a pronounced gap, we decrease the FLL DC energy by \SI{2.0}{\electronvolt}. Note that it has been argued that very often the FLL-DC is too high~\cite{HauleExactDC}. A reduction of the DC can also be accomplished by adjusting $U$ in the FLL formula~\cite{Millis5BandSrMnO3,ParkMillis_UprimeDC}. While we find that the \tg occupation is not much affected by the DC (similar to Fig.~\ref{fig:3BandDP_DC}), its effect on the \eg occupation is much stronger. Without a DC-shift, we find $\langle n_{e_{g} 0\sigma} \rangle \approx 0.30$, while for a shift by \SI{-2.0}{\electronvolt} the occupation is $\langle n_{e_{g} 0\sigma} \rangle \approx 0.19$ compared to $\langle n_{e_{g} 0\sigma} \rangle \approx 0.28$ in the DFT calculation. 
\\
Fig.~\ref{fig:5Band_DP} shows the spectral function of the full 5-band $d$-$dp$ calculation with adjusted DC as well as the respective spectral 
function obtained by a DMFT calculation using CTQMC and performing the analytic continuation for the impurity Green's function. Overall, the FTPS spectrum is in good agreement with the CTQMC+MaxEnt result. However, FTPS provides a much better energy resolution at high energies, which is especially apparent from the pronounced peak structure in the \eg spectrum.
\\
 Instead of calculating the real-frequency spectrum of the impurity Green's function as in Fig.~\ref{fig:5Band_DP}, one can use the analytic continuation for the self-energy $\Sigma(\omega)$, and calculate the local Green's function of the lattice model directly for real frequencies. This way, the dispersion of the DFT band structure enters on the real frequency axis directly, which increases the resolution of the CTQMC result, but is suboptimal for FTPS~\footnote{To calculate the self energy, we use the bath parameters to calculate the non-interacting Green's function. To avoid finite size effects, we need to use a larger broadening $\eta_{FT}$.}. The resulting spectral function for the \eg orbitals is shown in Fig.~\ref{fig:5Band_DP_Gloc}. As expected, we find that some features shown by the FTPS solver are now also present in CTQMC, but the $J$-multiplet in the unoccupied part of the spectrum still cannot be resolved. To calculate the FTPS self-energy, used to obtain the spectrum in Fig.~\ref{fig:5Band_DP_Gloc}, a broadening of $\eta_{FT} = 0.25$ was used, which explains the difference to Fig~\ref{fig:5Band_DP}.
 \\
 From these comparisons we also see that the sharp, step-like shape of the \eg spectrum at $E_F$ is not an artifact of the FTPS solver. We note that for the 5-band calculation presented in Fig.~\ref{fig:5Band_DP}, FTPS (720 CPU-h) and CTQMC (600 CPU-h) need similar computational effort for one DMFT iteration~\footnote{CTQMC used $128\cdot 10^6$ measurements and the calculations were performed on the same processors: Intel Xeon E5-2650v2, 2.6 GHz with 8 cores.}.
\\
The unoccupied part of the total spectrum (sum of the \eg and \tg spectra shown in the bottom plot of Fig.~\ref{fig:5Band_DP}) consists of a three peak structure with alternating \eg-\tg-\eg character, which is much more 
pronounced than in the 5-band $d$-only calculation (Fig.~\ref{fig:5Band}).
Compared to the 3-band $d$-$dp$ model we find differences mainly in the occupied part of the \tg spectral function (Fig.~\ref{fig:Comp3to5Band_DP}). This is especially apparent in the lowest peak, which seems 
to be shifted from \SI{-9.0}{\electronvolt} to \SI{-13.0}{\electronvolt}. Although this high energy excitation is small, the FTPS solver can reliably resolve it.

\begin{figure}[t]
   \centering
   \includegraphics{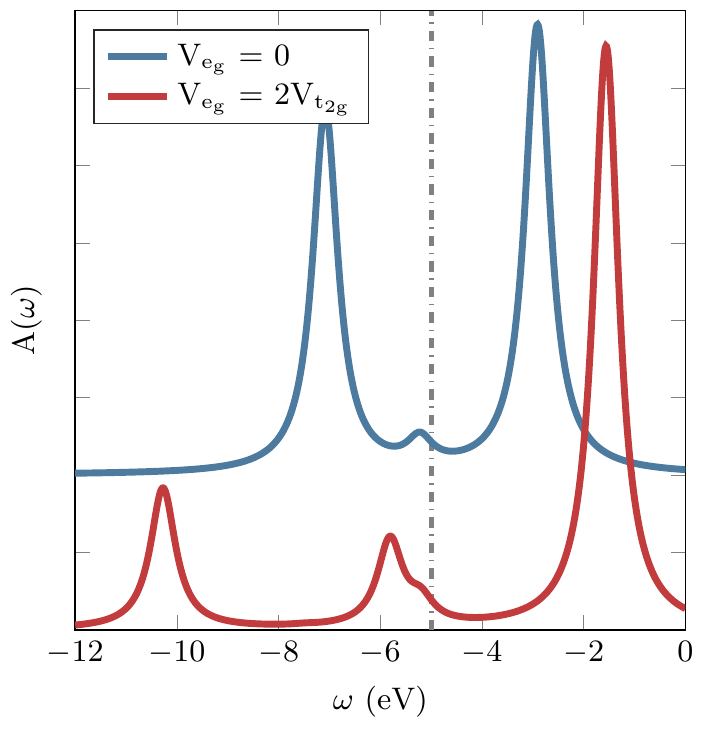}
   \caption{Effect of the \eg hybridizations on the \tg spectrum of the toy 
model (Eq.~\ref{eq:toymodel2}). Parameters (in eV): $U=10.0$, 
$J=U/10$, $E_{t_{2g}} = -U/2$, $E_{e_{g}} = -U/2+1.0$, 
$\epsilon_{t_{2g}} = \epsilon_{e_{g}} = -5.0$ and $V_{t_{2g}} = 1.5$. The grey dashed dotted line shows the bath energy 
levels. All spectra have been broadened by $\eta_{FT}=\SI{0.2}{\electronvolt}$.
}
   \label{fig:AIM_2orb1site}
\end{figure}

The differences in the position of this peak are again similar to the behavior of a toy model. Here we use a 
two-orbital AIM with a single bath site for each orbital:
\begin{align}
 H =& H_\text{int} + \sum_{m \in ( t_{2g}, e_g) } E_{m} n_{0,m} + \nonumber\\
  &\sum_{\sigma} V_m (c_{0,m,\sigma}^\dag 
c_{1,m,\sigma} + h.c.) + \epsilon_m n_{1,m,\sigma}\label{eq:toymodel2}.
\end{align}
For the interaction $H_{int}$ we choose the Kanamori Hamiltonian. As before, we use a single bath site for each orbital 
to mimic the effect of the HESW. We are interested in the influence of the hybridizations of 
\eg and O-$2p$ on the \tg spectral function. In Fig.~\ref{fig:AIM_2orb1site}, we compare the spectrum 
without \eg-HESW states ($V_{e_g}=0$) with the one obtained from $V_{e_g} = 2V_{t_{2g} 
}$\footnote{In the full 5-band calculation, the \eg bath spectral function is much larger than the one for the 
\tg orbitals in the energy region of the oxygen bands, which we mimic by a factor of 2 in $V_{e_g}$.}. 
Although one would expect the \eg hybridization to only have a minor influence on 
the \tg spectrum, we observe a rather surprising behavior. The additional 
hybridization leads to a stronger repulsion of the lowest energy peak from the bath energy, qualitatively explaining 
the shift from  \SI{-9.0}{\electronvolt} to \SI{-13.0}{\electronvolt} in Fig.~\ref{fig:Comp3to5Band_DP}. 
\\
Additionally, this toy model provides an explanation for the necessary adjustment of the DC correction in the 5-band 
calculation: The peak highest in energy in Fig.~\ref{fig:AIM_2orb1site} is repelled more strongly with the 
additional \eg hybridizations, therefore the gap decreases. If we would want to obtain a similar \tg gap as with 
$V_{e_g}=0$, the interaction in the toy model would need to be increased to $U \approx \SI{20}{\electronvolt}$ (keeping 
$J=U/10$). Since this is unphysical, the only other option is to shift the bath site energies of the toy model. In the 
DMFT calculation this corresponds to a shift in the DC correction, effectively shifting the HESW to lower energies. 
This behavior can be observed in Fig.~\ref{fig:Comp3to5Band_DP}, where we compare the spectra of the 3- and 5-band 
$d$-$dp$ models. The onset of the lower-Hubbard-band/HESW complex is exactly at the same position in both spectra, 
although the DC shift differs by \SI{2.0}{\electronvolt}.

\begin{figure}[t]
   \centering
   \includegraphics{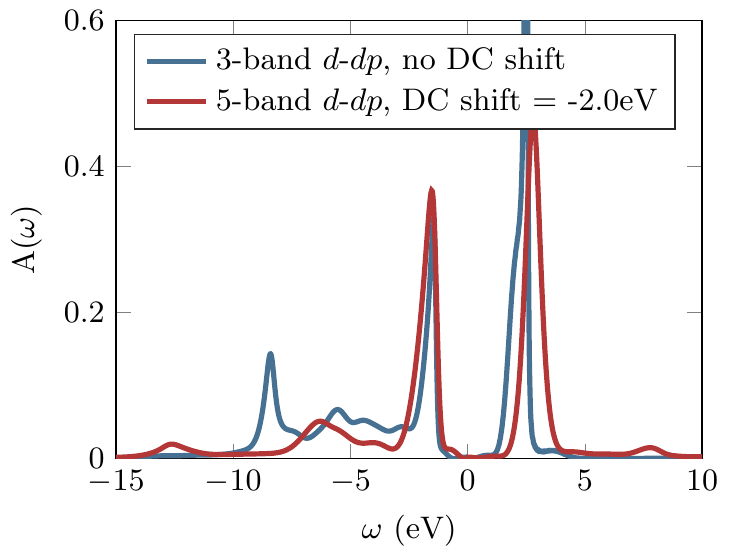}
   \caption{Comparison of the \tg spectral functions of the 3-band $d$-$dp$ and 5-band $d$-$dp$
calculations at $U=\SI{6.0}{\electronvolt}$ and $J=\SI{0.8}{\electronvolt}$, taken from Figs.~\ref{fig:3BandDP} 
and~\ref{fig:5Band_DP}. In the 5-band 
calculation we shifted the double counting by \SI{-2.0}{\electronvolt} to increase the gap. The influence of
the number of bands is most apparent in the high-energy features. The increased repulsion of the first peak of the 
lower-Hubbard-band/oxygen complex 
(Fig.~\ref{fig:AIM_2orb1site}) makes a shift in the DC necessary, if the single particle gap should remain the same.}
   \label{fig:Comp3to5Band_DP}
  \end{figure}

\begin{figure}[t]
   \centering
   \includegraphics{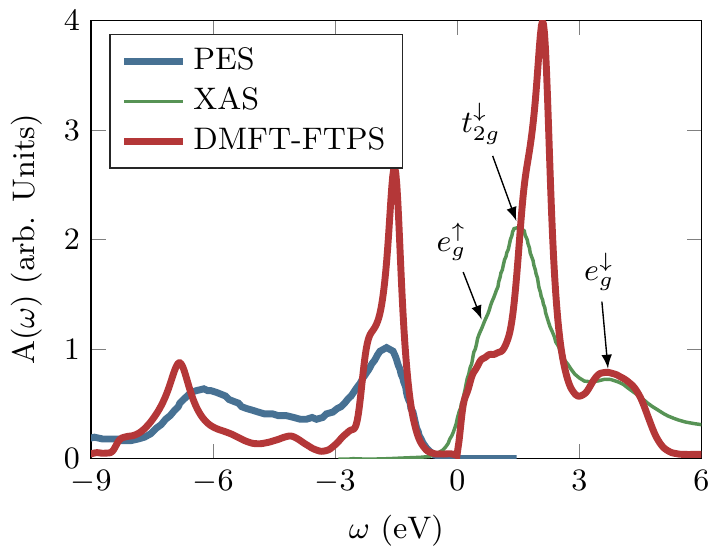}
   \caption{Photo Emission Spectroscopy (PES) and X-ray Absorption Spectroscopy 
(XAS) compared to the 5-band $d$-$dp$ DMFT-FTPS results ($U = \SI{5.0}{\electronvolt}$ and $J = 
\SI{0.6}{\electronvolt}$). The experimental curves are reproduced from Ref.~\onlinecite{KIM2010_EGT2GEG_minus13EV},
Fig.~5. We normalized the experimental curves to $\int_{-9}^0 A_{\text{PES}}d\omega = \int_{-9}^0 A_{\text{FTPS}} 
d\omega$  and $\int_{0}^6 A_{\text{XAS}}d\omega = \int_{0}^6 A_{\text{FTPS}} 
d\omega$. FTPS as well as the experiments show a 3-peak structure of alternating \eg - \tg - \eg character in the 
unoccupied part of the spectrum (indicated by arrows). For the arrow labels we adopted the notation of 
Ref.~\onlinecite{KIM2010_EGT2GEG_minus13EV}, where $e_g^\uparrow$ means an excitation into the \eg spectrum with 
majority spin, while $t_{2g}^\downarrow$ and $e_g^\downarrow$ are excitations into the \tg and \eg spectrum with
minority spin (see also Eq.~\ref{eq:egMinorityMajorityExcitation}).}
   \label{fig:CompToExp}
\end{figure}

\section{Comparison to experiment}\label{sec:ComparisonToExp}
Equipped with a good understanding of the \mbox{model-dependent} effects on the spectral function, we are finally in a 
position to compare our results to experiments. Several studies concluded that the unoccupied part of the spectrum 
consists of three peaks with alternating \eg - \tg - \eg character~\cite{KANG2008_EGT2GEG,Saitoh_EGEGT2G_minus13,
KIM2010_EGT2GEG_minus13EV}. As we have shown, with DMFT+FTPS we are able to resolve such a structure when including the \eg states as correlated orbitals in a genuine 5-band model. Additionally, we need to choose the energy window, i.e., 
whether the HESW should be included in the construction of the projective Wannier functions. The nature of the 
insulating state (Mott or charge transfer) has been debated in the
literature~\cite{Chainani1993,Saitoh_EGEGT2G_minus13,ABBATE_1992_T2GEGT2G,Millis5BandSrMnO3}, but  it is 
likely that \ce{SrMnO3} falls in an intermediate regime where a clear distinction is difficult. In the present work we 
have come to the same conclusion. This implies that the lower Hubbard band and the O-$2p$ bands are not separated in 
energy, which favors the use of a $d$-$dp$ model. We therefore conclude that a 
5-band $d$-$dp$ model is necessary to fully capture the low-energy physics of \ce{SrMnO3}. 
\\
Having decided on the model for the correlated subspace, we still need to determine the interaction parameters 
$U$ and $J$ as well as the DC. To do so we use PES and XAS data for the Mn-$3d$ orbitals obtained by Kim 
\emph{et al.}~\cite{KIM2010_EGT2GEG_minus13EV} and compare to our total impurity spectrum 
($6A_{t_{2g}}\left(\omega\right) + 4A_{e_g}\left(\omega\right)$ from Fig.~\ref{fig:5Band_DP}). According to 
Ref.~\onlinecite{KIM2010_EGT2GEG_minus13EV}, the XAS (PES) spectrum can be considered to represent the 
unoccupied (occupied) Mn-$3d$ spectrum. In the measured spectrum the chemical potential is in the middle of the gap. 
In all our calculations, the chemical potential is determined by the onset of the unoccupied \eg spectrum. However, the 
absolute position in energy is not exactly known in XAS~\cite{KangEMAIL}. Our calculation is in good agreement with the 
experiment when we use a rigid shift of the XAS spectrum by
\SI{0.8}{\electronvolt} to lower energies. Additionally, we deduce from the peak positions in the experiment that the interaction parameters used 
for the calculations presented in Fig.~\ref{fig:5Band_DP} are too high. The separation of the two \eg peaks ($\sim J$) and also the relative position of the \tg upper Hubbard band is different than in the experiment. Therefore, we decrease the interaction parameters
to $U = \SI{5.0}{\electronvolt}$ and $J = \SI{0.6}{\electronvolt}$ but keep the static shift of the FLL DC by 
\SI{-2.0}{\electronvolt}. Note that these parameters are similar to the ones used in other DFT+DMFT studies on \ce{SrMnO3}~\cite{Millis5BandSrMnO3,Millis5BandSrMnO3_2}. 
\\
The resulting spectral function for the new set of parameters is compared to the experimental spectrum in Fig.~\ref{fig:CompToExp}. 
Notably, the bandwidths of both, the unoccupied and the occupied spectrum, agree very well with the experiment. The 
unoccupied part of the experimental spectrum (XAS) shows that the first \eg peak is just a shoulder of the \tg upper 
Hubbard band, and that the separation of the two \eg peaks is about \SI{3.2}{\electronvolt}, which is in agreement 
with our result. Since this separation is proportional to the Hund's coupling, we conclude that $J\approx$ 
\SI{0.6}{\electronvolt} for this compound. The \tg upper Hubbard band at \SI{2.0}{\electronvolt} is still slightly 
too high in energy. \\
The experiment also shows a lower-Hubbard-band/oxygen complex with two main peaks at about 
\SI{-6.0}{\electronvolt} and \SI{-2.0}{\electronvolt}. As discussed in the previous sections (bottom plot of 
Fig.~\ref{fig:5Band_DP}), our results identify the first peak at \SI{-2.0}{\electronvolt} to have mainly \tg character 
and to correspond to the largest part of the split lower Hubbard band, whereas the second peak at 
\SI{-6.0}{\electronvolt} has both \eg and \tg 
character and stems from the hybridizations with the oxygen bands. We note that the region between these two peaks has larger spectral weight in the experiment than in our calculations. Importantly, no prominent spectral features are observed in the experiment 
around \SI{-8.0}{\electronvolt}, strengthening our conclusion that the 3-band $d$-$dp$ model is not sufficient to 
describe the experiment (see also Fig.~\ref{fig:Comp3to5Band_DP}).

\section{Conclusions}
We have studied the influence of the choice of the correlated subspace, i.e.\, the number of 
bands and the energy window, on the DFT+DMFT result for the strongly correlated compound
\ce{SrMnO3}. For $d$-only models (neglecting $p$-$d$ hybridizations), we 
have shown that the empty \eg orbitals should be included in the correlated subspace because interactions with 
the half-filled \tg bands affect the spectrum, leading to a multiplet structure and a broadening of the \eg 
\mbox{DFT-DOS}. Including the Mn-$3d$/O-$2p$ hybridizations in a 3-band model for the \tg bands only, i.e., the 
3-band $d$-$dp$ model, we found a situation similar to avoided crossing, which leads to an interesting 
interplay of atomic physics (lower Hubbard band) and Mn-$d$/O-$p$ hybridizations.
In \ce{SrMnO3}, the lower Hubbard band hybridizes with the \tg Wannier-weight on the oxygen bands, giving rise to a 
spectrum that can be approximated by three peaks. This result provides new perspectives on an intermediate regime, 
where both Mott and charge transfer physics are found. 
By performing a 5-band calculation including the $p$-$d$ hybridization, we investigated the effect of the \eg 
hybridization on the \tg spectrum. The splitting due to avoided crossing is heavily increased, which strongly affects 
the 3-peak structure and also decreases the gap. Equipped with a good understanding of the different correlated 
subspaces and the effects of the model parameters ($U$, $J$, DC) we were able to obtain a spectral function in good agreement with experimental data. We conclude that the choice of a suitable model for the correlated subspace is important, since the inclusion of both the O-$2p$ hybridizations and the \eg states is essential for a correct description of the observed spectral function in \ce{SrMnO3}. 
\\
Finally, we would also like to stress that we have shown that FPTS is a viable real-time impurity solver for real material calculations with five bands.

\section*{Acknowledgments}
The authors acknowledge financial support by the Austrian Science Fund
(FWF) through SFB ViCoM F41 (P04 and P03), through project P26220, and through the START program Y746, as well as by NAWI-Graz. 
This research was supported in part by the National Science Foundation under Grant No. NSF PHY-1125915. We thank J.-S. 
Kang for the permission to reproduce data and for helpful discussions. We are grateful for stimulating discussions 
with J. Mravlje, F. Maislinger and G. Kraberger. The computational resources have been provided by the Vienna
Scientific Cluster (VSC). All calculations involving tensor networks
were performed using the ITensor library~\cite{ITensor}.

 \bibliography{paper.bib}
\end{document}